# Relationship between Perceived Maneuverability and Involuntary Eye Movements under Systematically Varied Time Constants of Ride-on Machinery

Muhammad Akmal Bin Mohammed Zaffir[1], Daisuke Sakai[1], Yuki Sato[2], and Takahiro Wada[1]

[1] Division of Information Science, Nara Institute of Science and Technology, 8916-5, Takayama-cho, Ikoma, Nara, 630-0192, Japan

[2] Faculty of Applied Science and Engineering, Ibaraki University, 4-12-1, Nakanarusawa-cho, Hitachi, Ibaraki, 316-8511, Japan

**Corresponding Author:** Takahiro Wada, E-mail: t.wada@is.naist.jp, **ORCID:** 0000-0002-4518-8903

Note: This is the originally submitted version of the manuscript titled

"Relationship between Perceived Maneuverability and Involuntary Eye Movements under Systematically Varied Time Constants of Ride-on Machinery."

The final version, which incorporates revisions made during peer review, has been published under the title

"Relationship between perceived maneuverability and compensatory eye movements under systematically varied time constants of ride-on machinery" in Experimental Brain Research.

Citation:

Muhammad Akmal Bin Mohammed Zaffir, Daisuke Sakai, Yuki Sato, Takahiro Wada,

"Relationship between perceived maneuverability and compensatory eye movements under systematically varied time constants of ride-on machinery," Experimental Brain Research (2025).

https://doi.org/10.1007/s00221-025-07135-3

**Abstract:**

Studies suggest that involuntary eye movements exhibit greater stability during active motion compared to passive motion, and this effect may also apply to the operation of ride-on machinery. Moreover, a study suggested that experimentally manipulating the sense of agency (SoA) by introducing delays may influence the stability of involuntary eye movements. Although a preliminary investigation examined involuntary eye movements and perceived maneuverability under two distinct machine dynamics with preserved SoA, it remains unclear how systematic variations in motion dynamics influence these factors. Therefore, the purpose of the present research was to investigate whether systematic variations in the dynamic properties of a ride-on machine, where the perceived maneuverability is modulated, influence the accuracy of involuntary eye movements in human operators.

  Participants rode a yaw-rotational platform whose time constant from joystick input to motor torque of a rotational machine was systematically manipulated. During the operation, eye movements were recorded while participants fixated on a visual target. After each condition, participants provided subjective ratings of maneuverability and cognitive load. As the platform's time constant increased, the perceived maneuverability scores decreased while the cognitive loads increased. Concurrently, involuntary eye movement accuracy decreased. Moderate to weak positive correlations emerged between the perceived maneuverability scores and the eye movement gain and accuracy, while a weak negative correlation was found with cognitive load. These findings suggest that subjective feelings of control and gaze stabilization are closely intertwined under varying time constants of machine dynamics. By elucidating the connection between perceived maneuverability and involuntary eye movement accuracy, this study could provide insights for designing ride-on machinery with high maneuverability.

**Introduction**

In this study, we aim to understand perceived maneuverability during machine operation from the perspective of human motor control. Perceived maneuverability depends not only on the physical properties of a machine but also on the operator's ability to predict and adapt to its movements. As tasks involving various machines and robots continue to expand, understanding the mechanisms underlying this perception is increasingly important. However, most previous research studies have assessed maneuverability primarily based on task performance and subjective workload measures, as seen in various applications such as automobiles(Nakayama et al., 1999), construction machinery (Akyeampong et al., 2014; Okawa et al., 2020). We suggest that incorporating insights from human motor control may lead to a deeper understanding of machine maneuverability. The present study focuses on "ride-on machinery," such as automobiles or construction equipment, where the operator's body moves in response to the machine's motion, which itself is driven by the operator's control inputs.

During both one's own body movements and those performed while operating ride-on machinery, the control of eye movements is essential for maintaining an accurate perception–action loop. For example, in situations where a visual target is present, involuntary eye movements including the vestibulo-ocular reflex (VOR)—which occurs in the opposite direction of head movement—help the operator maintain accurate gaze on the fixed target. It has been reported that VOR can be influenced by factors such as the presence or absence of a fixation target (Demer et al., 1993; Jell et al., 1988; Paige et al., 1998), the distance to that target (Paige et al., 1998), and changes in gaze direction（Angelaki & Cullen, 2008; Medendorp et al., 2000）. Some research studies have demonstrated that cognitive load affects accuracy of involuntary eye movements(Kono et al., 2019, 2023; Le et al., 2020; Obinata et al., 2008).

Furthermore, previous studies suggest that the volition behind head or body movements influences VOR gain (defined as the ratio of eye velocity to head velocity), with active head or body motion differing from passive motion (Demer et al., 1993; Jell et al., 1988), although some studies challenge the extent of these differences（Furman & Durrant, 1998; Hanson & Goebel, 1998; Schubert & Migliaccio, 2016）. These effects of the presence or absence of movement intention on involuntary eye movements can be interpreted as the influence of motion prediction, as follows. When performing active movements, humans utilize the efference copies and internal representations (internal models) of their body and sensory apparatus to enhance the accuracy of prediction and control of their own motion (Kawato, 1999; Merfeld et al., 1999; Wolpert et al., 1995). This predictive information is considered to play an important role in maintaining stable gaze（Cullen & Roy, 2004）. Furthermore, computational models have been developed to describe how humans process vestibular signals, including 'GIF resolution'—that is, decomposing gravito-inertial acceleration (GIF) into its gravitational and inertial components—and how this process is integrated into the VOR（Merfeld & Zupan, 2002; Uefune et al., 2016）. In addition, when operating ride-on machinery, it is likely that knowledge of the machine's dynamic properties—namely, how the vehicle responds to control inputs—is also leveraged as part of this internal model. Indeed, besides the reported changes in VOR gain between active and passive conditions, one study (Uefune et al., 2018) suggests that VOR accuracy improves when operating a machine actively, compared with passive movement. Another line of work (Sato et al., 2020) suggests that when there is a delay between an operator's input and the resulting ride-on machine (or body) motion, or when the motion direction is reversed (i.e. the gain becomes negative), the Sense of Agency (SoA), which refers to the feeling of controlling one's own actions (Haggard & Chambon, 2012) tends to decrease. Their study further suggested that under such conditions, the accuracy of involuntary eye movements may also decrease. Additionally, a preliminary study suggests that once operators have adapted to a given set of machine dynamics, abruptly altering those dynamics can degrade the performance of involuntary eye movements (Sato et al., 2021). These findings point to the possibility that the aspect of a machine's subjective maneuverability arising from the ease of predicting its movement may, in turn, be related to the accuracy of involuntary eye movements. In other words, greater accuracy in motion prediction could lead to more stable involuntary eye responses, such as improved fixation accuracy. However, to our knowledge, no study has directly examined how subjective maneuverability and the accuracy of involuntary eye movements are related in more gradual and systematic variations in the dynamic properties of the operated machine.

In the present study, the machine's dynamic properties were systematically varied under multiple time constants (delays), and participants were instructed to operate the machine in each condition, enabling us to evaluate both subjective maneuverability and fixation accuracy of the involuntary eye movements. We hypothesized that the accuracy of the eye movements is higher in environments where perceived maneuverability is higher. Specifically, participants rode a machine capable of yaw rotation controlled via a joystick. While they operated the machine, we recorded their eye movements; afterward, participants provided subjective ratings of maneuverability. Then the relationship between these maneuverability ratings and eye movement accuracy was investigated.

The primary contribution of the present study is to provide empirical evidence linking changes in the perceived maneuverability of ride-on machinery—induced by systematic variations in the machine's time constant of the dynamic properties—to the performance of involuntary eye movements. The remainder of this paper is organized as follows. Section 2 describes the experimental setup and procedures, including details of the apparatus and participants. Section 3 presents the results of our analyses, focusing on the accuracy of involuntary eye movements and subjective maneuverability ratings. In Section 4, we discuss the theoretical and practical implications of our findings and consider potential directions for future research. Finally, Section 5 summarizes the conclusions drawn from this study.

**Method**

**2.1 Apparatus**

Figure.1 shows the yaw rotational platform (yawing chair) used in this study. The chair can rotate from -180 deg to 180 degrees around the Earth-vertical axis, controlled by a joystick located on the right-hand side. By tilting the joystick left or right, the chair rotates in the corresponding direction around the vertical axis. Figure. 2 shows the overall system architecture. The system includes a digital signal processor (DSP) board (sBOX2, MTT Corporation), which connects the chair motor controller. The angular acceleration $\ddot{\theta}$ [rad/s$^2$] of the motor (chair) is proportional to voltage $v$, supplied to the motor controller. Additionally, the voltage $v$ was determined based on the joystick angle $\delta(t)$ [deg] using first-order lag dynamics. As a result, the relationship between the joystick angle $\delta$ and the motor's angular acceleration $\ddot{\theta}$ can be expressed by the first-order lag dynamics shown in Eq. (1):

$$\ddot{\theta} = \frac{K}{Ts+1}\delta \tag{1}$$

where the time constant $T$ [s] was configured by the DSP while the gain $K$ was determined by the moment of inertia of the chair and the settings of the DSP and motor controller. The current chair angle $\theta(t)$, measured by the rotary encoder, was sent to a visual monitor to display an indicator. The visual display was placed approximately 1 meter in front of the chair. Figure 2-(a) illustrates the structure of the indicator. The outer and inner diameters of the annulus corresponded to approximately 4.0 and 1.7 degrees of visual angle, respectively, while the fixation point at its center had a visual angle of approximately 0.30 degrees. A white line rotating around a fixation point, represents the current chair angle $\theta(t)$. Two blue annular sectors, representing the target areas for chair angular position, were displayed to indicate the range within which participants were expected to move the chair. The remaining annular area serves as the control state indicator, which changes color according to the chair angle $\theta(t)$ as follows (see also Fig. 2(b)):

1.  Green indicates $0\deg > \theta > -25\deg$ or $0\deg < \theta < 25\deg$, indicating the chair has not reached the target area.
2.  Blue indicates $-25\deg \geq \theta \geq -35\deg$ or $25\deg \leq \theta \leq 35\deg$, indicating the chair is within the target area.
3.  Red indicates $\theta < -35\deg$ or $\theta > 35\deg$, indicating the chair exceeds the target area.

A wearable eye-tracking device (Tobii Pro Glasses 3, Tobii Technology K.K.) was used to measure the operator's eye direction and head angular velocity. Participants wore a neck stabilization device throughout the experiment to minimize head movement relative to the trunk and chair. Black curtains were placed around the yawing chair and the indicator to reduce visual distractions.

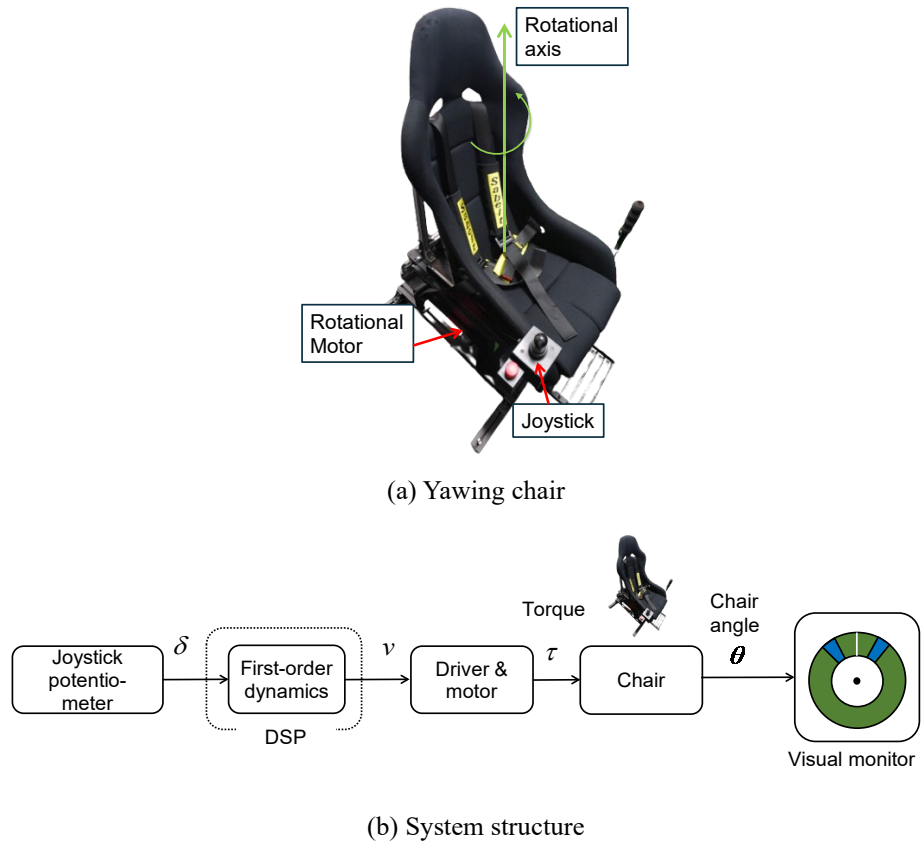

(a) Yawing chair

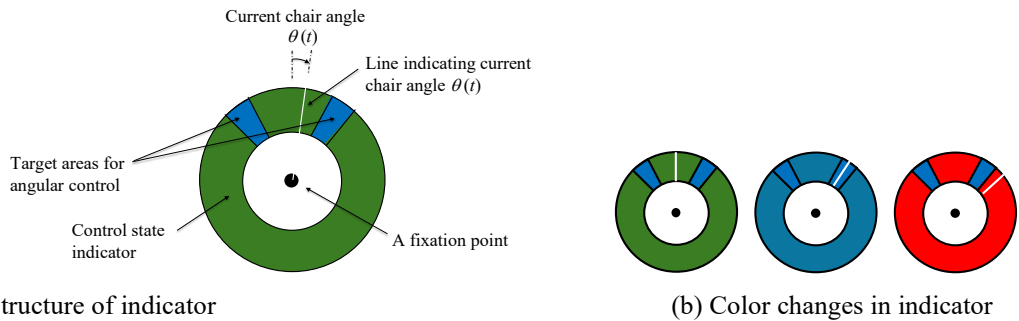

(b) System structure

**Fig. 1 Experimental apparatus**

(a) Structure of indicator

(b) Color changes in indicator

**Fig. 2:** An indicator displayed on the visual monitor. (a) Structure of the indicator: A white line, rotating around a fixation point, represents the current chair angle $\theta(t)$, with two blue annular sectors indicating the target areas for angular control. The remaining annular area represents the control state indicator, where the color changed according to the current chair angle $\theta(t)$. (b) Color changes in the control state indicator: Green indicates that $\theta$ is below the target area; Blue indicates that $\theta$ is within the target area; Red indicates $\theta$ exceeds the target area.

## 2.2 Experimental Design

The dynamics of the yawing chair were changed by changing the time constant $T$ in Eq.(1) at four levels

($T$ = 0.0, 0.1, 0.2, and 0.4 s), as an experimental factor. The chair dynamics condition was treated as a within-subject factor. Each participant experienced all four levels of the dynamics change in a single day. The order of conditions was counterbalanced to eliminate order effects. This experiment was approved by the Research Ethics Board of Nara Institute of Science and Technology (No. 2023-I-19).

## 2.3 Task

The participants were instructed to sit on the chair with the seat belt fastened and to wear a neck stabilization device. They were required to gaze at the fixation point, which was the center of the indicator on the visual display, throughout the experimental trials. Since the chair's rotation caused the head to rotate by nearly the same amount due to neck stabilization, fixation on the point had to be achieved solely through eye movements relative to head motion.

As shown in Fig. 3, the chair position perpendicular to the monitor was defined as 0 degrees, and two target areas were set within the ranges of -25 to -35 degrees and 25 to 35 degrees, each spanning 10 degrees. Participants were instructed to control the chair using a joystick, synchronizing their movements with auditory metronome cues, to move back and forth between the left and right target areas. The metronome frequency was set to 0.75 Hz. As they were required to maintain fixation on the designated point, they had to rely on the indicator displayed on the visual monitor to perceive the current chair position and the target areas. Each participant experienced four experimental trials with different dynamics settings described in Section 2.4. Each condition lasted 90 seconds.

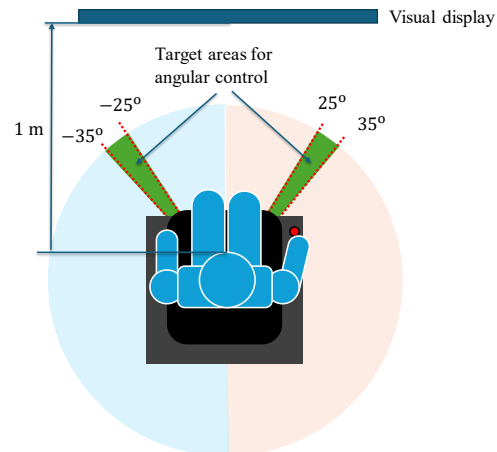

**Fig. 3: Top view of experimental setup**

## 2.4 Procedure

The experiment was conducted on a single day for each participant using the following procedure. First, the details of the experiment, including its research objectives and procedures, were thoroughly explained to the participants. They were informed that they could withdraw at any time, during or after the experiment.

All participants provided written consent.

To begin, the participants sat in the chair, wore Tobii Pro Glasses 3, fastened their seatbelts, and wore a neck stabilization device to minimize the relative motion between the chair and their head. Subsequently, they were given instructions on the task again.

Then, participants completed the task described in Section 2.3 under four experimental conditions in the order assigned to them. For each condition, a practice trial lasting 20 seconds was followed by a 90-second measurement trial. After both trials in each condition, participants completed a questionnaire regarding their perceived maneuverability under the given dynamic condition. This process was repeated for all four conditions. Details of the questionnaire are provided in Section 2.6.3.

**2.5 Participants**

This study included 24 participants (15 males and 9 females) with an average age of 24.6 years (SD = 3.5). None of them had been diagnosed with photosensitive epilepsy, vestibular disorders, or balance disorders. Additionally, they were also not taking any medication that induces drowsiness. Participants who wore glasses were instructed to wear contact lenses, as the eye-tracking device was designed in the form of eyeglasses.

**2.6 Data Collection and Analysis Method**

2.6.1 Data collection and pre-processing

Head angular velocities $\omega_{head}$ $(\in R^3)$ were measured at 100 Hz with the IMU sensor embedded in the Tobii Pro Glasses 3 and downsampled to 50 Hz. Normalized gaze directions $d$ $(\in R^3)$ were measured for the left and right eyes at 50 Hz by the Tobii Pro Glasses 3. The gaze direction vector was processed with a zero-phase low-pass filter with a cutoff frequency of 1 Hz. The eye angular velocity $\omega_{eye}$ $(\in R^3)$ was calculated using the eye direction vector $d$ and its time derivative approximated by the second-order central difference.

Numerous missing values were observed in the eye direction dataset. To address this, if the duration of consecutively missing values exceeded a certain time threshold, we discarded the data segments containing the missing values. Otherwise, we applied linear interpolation. In the present study, a time threshold of 0.6 s was applied.

2.6.2 Eye movement metrics

The gain $G$ is defined as in Eq. (4):

$$\omega_{eye}^{yaw} = -G\omega_{head}^{yaw} \tag{4}$$

where $\omega_{eye}^{yaw}$ and $\omega_{head}^{yaw}$ denote scalars representing the angular velocities of the eyes and head, respectively. $G$ was identified using the least squares method for each sliding time window of 2.76 s, which corresponds to one cycle of the chair's rotational motion.

Eye movement error $e_{eye}$ was defined as in Eq. (5).

$$e_{eye} := \omega_{eye}^{yaw} + \omega_{head}^{yaw} \qquad (5)$$

The root mean square (RMS) value of $e_{eye}$ was compared among conditions in the present study.

2.6.3 Subjective evaluation metrics

In this study, participants' perceived maneuverability was subjectively assessed using questionnaires, as summarized in Table 1. Q1 examined whether the yawing chair's motion initiation timing matched participants' expectations when moving the joystick. Q2 evaluated whether participants were able to stop the chair at the expected target timing and location during the task. Q3 assessed whether the chair's motion synchronized with the target angular positions and the metronome's cycle. Q4 examined whether the chair's movements aligned with participants' operational intentions. Q5 determined whether participants needed to adjust their operations to compensate for the machine's movements. The responses to the questionnaires were collected using a 7-point Likert scale, with 7 indicating "strongly agree," 1 indicating "strongly disagree," and 4 indicating "neither agree nor disagree." Additionally, participants' workload was evaluated using the Weighted Workload (WWL) scores from the Japanese version（Haga & Mizukami, 1996）of the NASA Task Load Index (NASA-TLX)（Hart & Staveland, 1988）.

Table 1: Questionnaires to subjectively evaluate participants' subjective maneuverability

| No. | Questions |
| --- | --- |
| Q1 | The timing at which the machine started moving was as expected |
| Q2 | The machine's motion could be stopped at the expected timing and location |
| Q3 | The chair could be operated in sync with the target angle and the metronome's cycle |
| Q4 | The machine's movements in response to my operation were as desired (or as preferred) |
| Q5 | There was no need to make an effort to adjust to the machine's motion |

2.6.4 Statistical analysis

For eye movement metrics, since either homoscedasticity or normality was not satisfied, the Friedman test was used to investigate the main effects of the chair dynamics condition. The Friedman test was also applied to subjective scores (questionnaires and NASA-TLX). When a significant effect was found in the test, post-hoc Wilcoxon signed-rank tests with Benjamini-Hochberg (BH) correction were performed.

We also performed correlation analysis using Pearson's correlation coefficients to investigate the relationship between VOR metrics and subjective scores.

3. Results

3.1 Results of eye movement metrics

Figure 4-(a) illustrates the eye movement gain for each condition. A Friedman test on the eye movement gain revealed a statistically significant main effect of the chair dynamics condition for both left ( $\chi^2(3) = 20.6$, $p = 0.000$ ) and right ( $\chi^2(3) = 22.7$, $p = 0.000$ ) eyes. Subsequent post-hoc pairwise

comparisons using the Wilcoxon signed-rank test with BH correction revealed the following results (see Table 2): Gain with $T = 0.0$ was significantly larger than that with $T = 0.1$ for both the left and right eyes. Gain with $T = 0.0$ was significantly larger than that with $T = 0.2$ for the left eye and marginally significantly larger for the right eye. Gain with $T = 0.0$ was significantly larger than that with $T = 0.4$ for both eyes. No significant difference was found between $T = 0.1$ and $T = 0.2$ for both eyes. Finally, gains with $T = 0.1$ and $T = 0.2$ were significantly larger than that with $T = 0.4$, for the right eye. Please refer to Table 2 for details.

**Table 2. Details of post-hoc pairwise tests for eye movement gain**

| Condition 1 | Condition 2 | Eyeside | z | P |
|---|---|---|---|---|
| $T = 0.0$ | $T = 0.1$ | Left | -5.29 | 0.047* |
| | | Right | -5.40 | 0.015* |
| $T = 0.0$ | $T = 0.2$ | Left | -5.61 | 0.003** |
| | | Right | -5.17 | 0.068† |
| $T = 0.0$ | $T = 0.4$ | Left | -5.87 | 0.000*** |
| | | Right | -5.79 | 0.000*** |
| $T = 0.1$ | $T = 0.2$ | Left | -5,87 | 0.665 |
| | | Right | -4.62 | 1.000 |
| $T = 0.1$ | $T = 0.4$ | Left | -4.77 | 0.103 |
| | | Right | -5.47 | 0.015* |
| $T = 0.2$ | $T = 0.4$ | Left | 5.20 | 0.067† |
| | | Right | -5.69 | 0.002** |

Fig. 4-(b) illustrates the RMSE of eye movement for each condition. A Friedman test on the RMSE revealed a statistically significant main effect of the chair dynamics condition for both left ( $\chi^2(3) = 11.7, p = 0.00843$ ) and right ( $\chi^2(3) = 10.3, p = 0.0161$ ) eyes. Subsequent post-hoc pairwise comparisons using the Wilcoxon signed-rank test with BH correction revealed the following results (see Table 3): The RMSE values with $T = 0.0$ were significantly smaller than those with $T = 0.2$ and $T = 0.4$ for both eyes. Additionally, RMSE with $T = 0.1$ was significantly smaller than that with $T = 0.2$ for the left eye and marginally significantly smaller for the right eye. No significant differences were found between $T = 0.0$ and $T = 0.1$, nor $T = 0.1$ and $T = 0.2$.

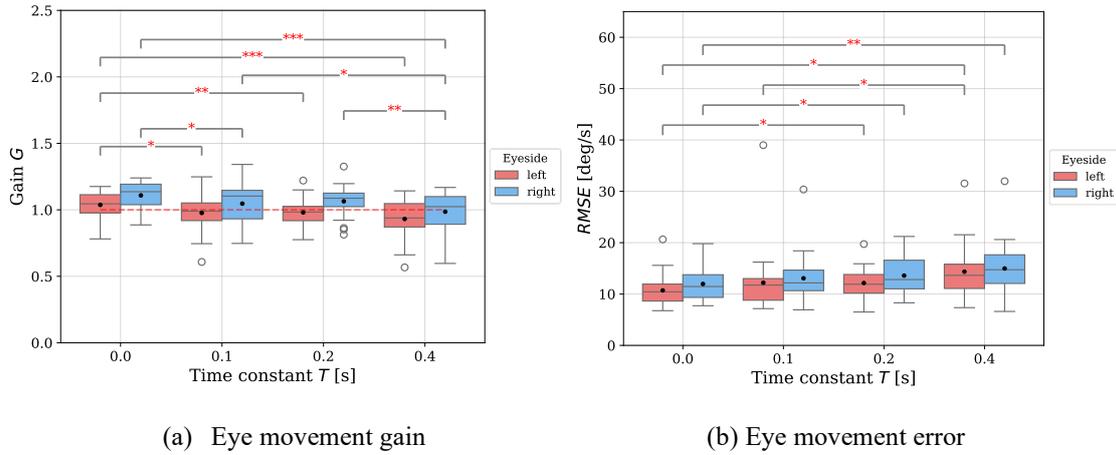

(a)  Eye movement gain

(b)  Eye movement error

**Fig. 4: Results of eye movement metrics**

Table 3. Details of post-hoc pairwise tests for RMSE of eye movement error

| Condition 1 | Condition 2 | Eyeside | z | P |
|---|---|---|---|---|
| $T = 0.0$ | $T = 0.1$ | Left | -5.02 | 0.181 |
| | | Right | -4,81 | 0.472 |
| $T = 0.0$ | $T = 0.2$ | Left | -5.40 | 0.027* |
| | | Right | -5.39 | 0.030* |
| $T = 0.0$ | $T = 0.4$ | Left | -5.53 | 0.013* |
| | | Right | -5.69 | 0.030** |
| $T = 0.1$ | $T = 0.2$ | Left | -5,06 | 0.200 |
| | | Right | -5.09 | 0.254 |
| $T = 0.1$ | $T = 0.4$ | Left | -5.42 | 0.033* |
| | | Right | -5.25 | 0.121 |
| $T = 0.2$ | $T = 0.4$ | Left | -5.28 | 0.072† |
| | | Right | -4.85 | 0.501 |

*: p < 0.05, **: p < 0.01, ***: p < 0.001

### 3.2  Results of subjective evaluations

Figure. 5 presents the results of the questionnaires on perceived maneuverability. Friedman tests of the questionnaire scores revealed a statistically significant main effect of the chair dynamics condition for all questionnaires (Q1: $\chi^2(3) = 32.5$, $p = 0.000$ ; Q2: $\chi^2(3) = 34.9$, $p = 0.000$ ; Q3: $\chi^2(3) = 36.1$, $p = 0.000$ ; Q4: $\chi^2(3) = 38.5$, $p = 0.000$ ; Q5: $\chi^2(3) = 21.7$, $p = 0.000$ ). Subsequent post-hoc pairwise comparisons revealed that participants perceived maneuverability to be significantly more difficult as the time constant increased by 0.2 s or more as shown in Table 4. Specifically, comparisons across all questions revealed that the scores were significantly higher for $T = 0.0$ than for $T = 0.2$, $T = 0.0$ than for $T = 0.4$, and $T = 0.1$ than for $T = 0.4$. Additionally, the scores for Q2 and Q3 were significantly higher with $T = 0.2$ than with $T = 0.4$. Furthermore, significant differences were observed between $T = 0.1$ and $T = 0.2$ for Q2–Q4.

Fig. 6 shows the result of the WWL score of the NASA-TLX. A Friedman test revealed a statistically significant main effect of the chair dynamics condition ($\chi^2(3) = 31.8$, $p = 0.000$). Subsequent post-hoc pairwise comparison revealed that WWL scores were significantly higher with $T = 0.0$ than with $T = 0.2$ ($z = -5.69$, $p = 0.001$) and $T = 0.4$ ($z = -5.96$, $p < 0.001$), with $T = 0.1$ than with $T = 0.2$ ($z = -5.37$, $p = 0.015$) and $T = 0.4$ ($z = -5.86$, $p < 0.001$), as well as with $T = 0.2$ than with $T = 0.4$ ($z = -5.38$, $p = 0.015$).

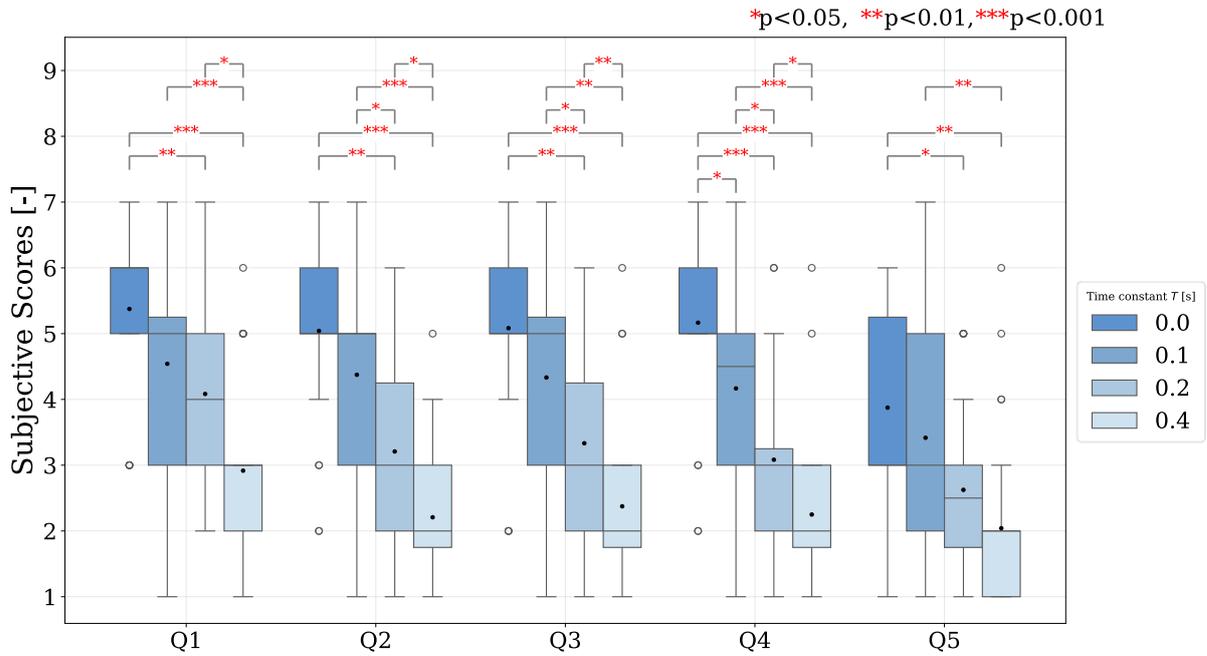

**Fig. 5 Subjective scores for perceived maneuverability**

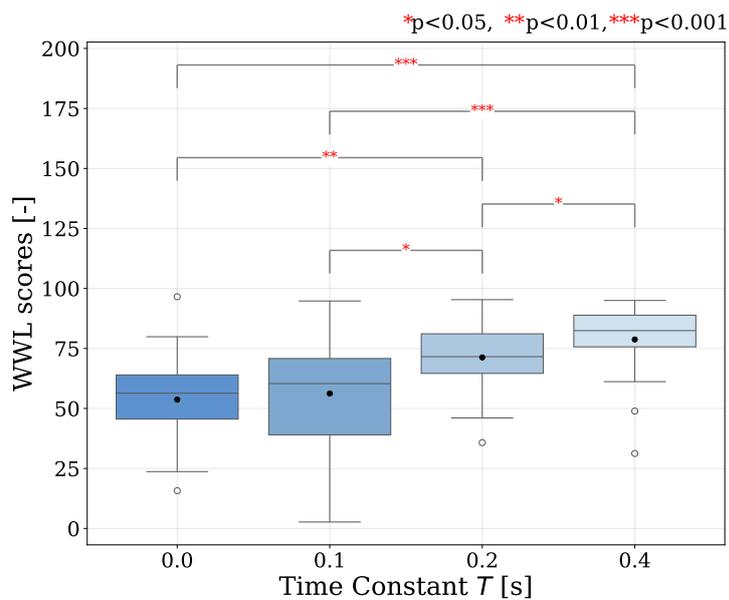

**Fig. 6 WWL scores**

Table 4. Post-hoc pairwise comparisons for questionnaire scores on perceived maneuverability

| Condition 1 | Condition 2 | Questions | Z | p |
|---|---|---|---|---|
| T = 0.0 | T = 0.1 | Q1 | -5.57 | 0.098† |
| | | Q2 | -5,36 | 0.235 |
| | | Q3 | -5.58 | 0.130 |
| | | Q4 | -5.65 | 0.044* |
| | | Q5 | -5.55 | 0.478 |
| T = 0.0 | T = 0.2 | Q1 | -5.88 | 0.005** |
| | | Q2 | -5.83 | 0.003** |
| | | Q3 | -5.82 | 0.004** |
| | | Q4 | -5.82 | 0.001** |
| | | Q5 | -5.72 | 0.033* |
| T = 0.0 | T = 0.4 | Q1 | -6.03 | 0.000*** |
| | | Q2 | -5.97 | 0.000*** |
| | | Q3 | -5.95 | 0.000*** |
| | | Q4 | -6.00 | 0.000*** |
| | | Q5 | -5.93 | 0.003** |
| T = 0.1 | T = 0.2 | Q1 | -5.33 | 0.173 |
| | | Q2 | -5.54 | 0.024* |
| | | Q3 | -5.48 | 0.041* |
| | | Q4 | -5.71 | 0.009** |
| | | Q5 | -5.49 | 0.142 |
| T = 0.1 | T = 0.4 | Q1 | -6.02 | 0.000*** |
| | | Q2 | -5.96 | 0.000*** |
| | | Q3 | -5.88 | 0.001** |
| | | Q4 | -6.02 | 0.000*** |
| | | Q5 | -5.99 | 0.003** |
| T = 0.2 | T = 0.4 | Q1 | -5.82 | 0.008* |
| | | Q2 | -5.78 | 0.021* |
| | | Q3 | -5.76 | 0.015* |
| | | Q4 | -5.53 | 0.067† |
| | | Q5 | -5.59 | 0.142 |

*: $p < 0.05$, **: $p < 0.01$, ***: $p < 0.001$

### 3.3 Correlation

Figure. 7-(a) and (b) show the correlation between participants' perceived maneuverability and the eye movement gain and error, respectively. Table 5, which summarizes the results of the correlation analysis for eye movement gain, revealed moderate to weak positive correlations with statistical significance between the standardized scores in the questionnaires and standardized eye movement gain. A statistically significant weak negative correlation was observed between the standardized WWL score and the standardized eye movement gain. Regarding the eye movement error metric, Table 6 showed weak negative correlations with statistical significance between the standardized scores in the questionnaires and standardized eye movement error, except for Q5, which reflects the effort required in machine operation. Additionally, a significant weak positive correlation was observed between the standardized WWL score and the standardized eye movement error.

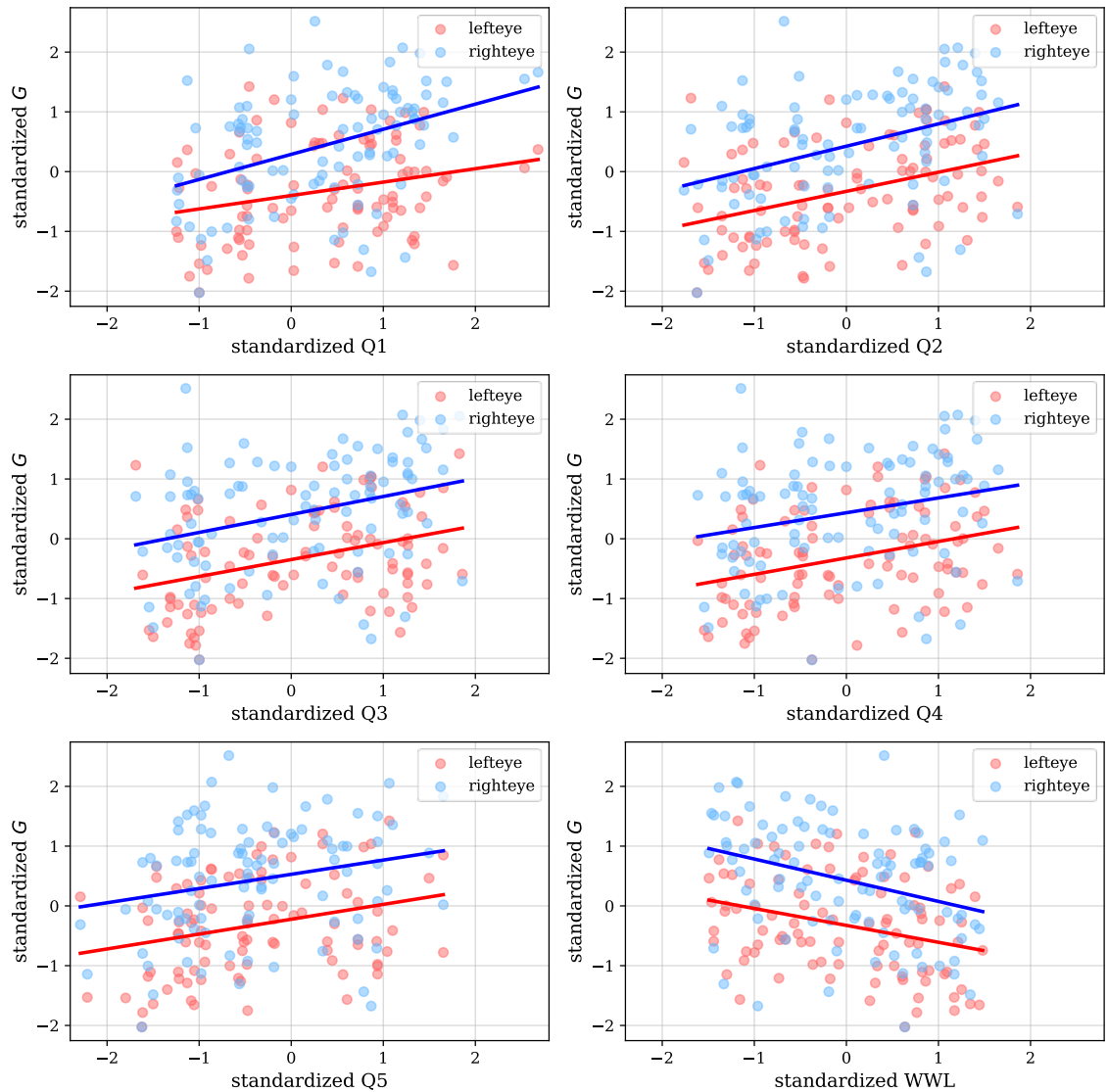

(a) Eye movement gain vs. perceived maneuverability

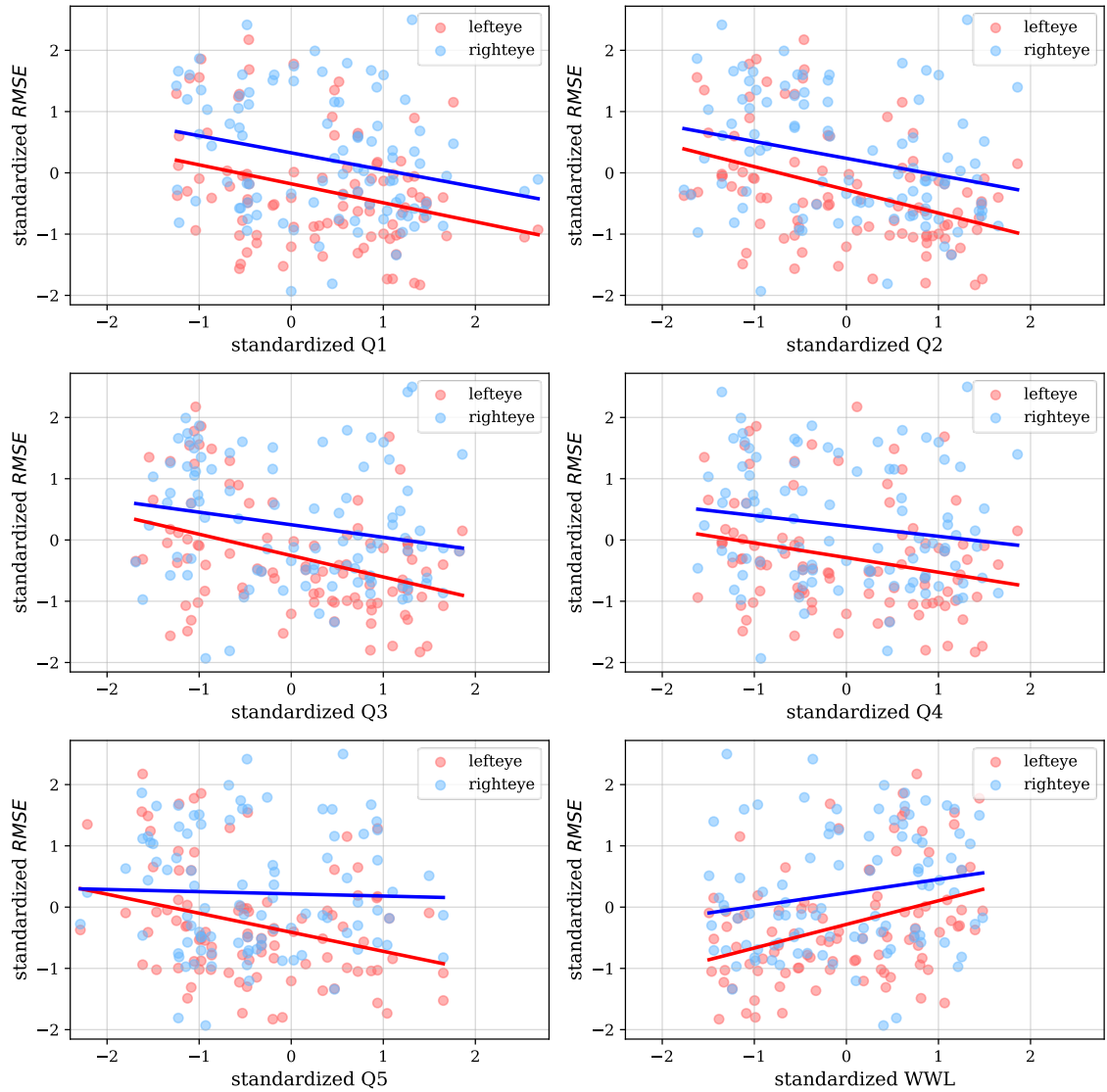

(b) Eye movement error vs. perceived maneuverability

**Fig. 7 Correlation between eye movement metrics and perceived maneuverability**

**Table 5: Correlation analysis between subjective evaluations and eye movement gain**

| Pair (standardized) | Eyeside | Pearson's correlation Coefficient    $r$ | $p$ |
|---|---|---|---|
| Q1-gain | Left | 0.25 | 0.014* |
|  | Right | 0.40 | 0.000*** |
| Q2-gain | Left | 0.39 | 0.000*** |
|  | Right | 0.39 | 0.000*** |
| Q3-gain | Left | 0.35 | 0.001** |
|  | Right | 0.32 | 0.002** |
| Q4-gain | Left | 0.32 | 0.002** |
|  | Right | 0.25 | 0.016* |
| Q5-gain | Left | 0.28 | 0.007** |
| | Right | 0.23 | 0.028* |
| WWW-gain | Left | -0.30 | 0.003** |
| | Right | -0.32 | 0.001** |

*: $p < 0.05$, **: $p < 0.01$, ***: $p < 0.001$

**Table 6: Correlation analysis on subjective evaluations and eye movement error**

| Pair (standardized) | Eyeside | Pearson's correlation Coefficient  $r$ | $p$ |
|---|---|---|---|
| Q1-RMSE | Left | -0.29 | 0.004** |
| | Right | -0.26 | 0.012* |
| Q2-RMSE | Left | -0.35 | 0.000*** |
| | Right | -0.29 | 0.005** |
| Q3-RMSE | Left | -0.36 | 0.000*** |
| | Right | -0.27 | 0.008** |
| Q4-RMSE | Left | -0.25 | 0.014* |
| | Right | -0.21 | 0.043* |
| Q5-RMSE | Left | -0.20 | 0.056 |
| | Right | -0.01 | 0.962 |
| WWW-RMSE | Left | 0.302 | 0.003** |
| | Right | 0.216 | 0.035* |

*: $p < 0.05$, **: $p < 0.01$, ***: $p < 0.001$

## 4. Discussion and conclusion

### 4.1 Summary of results

We investigated the relationship between the operator's subjective perception of maneuverability and involuntary eye movements by varying the yaw rotation dynamics of a ride-on machine through changes in the time constant from joystick angle input to chair acceleration output. As the time constants increased, the stability of involuntary eye movements decreased from the viewpoints of gain (Fig.4-(a) and Table 2) and errors (Fig.4-(b) and Table 3), while subjective maneuverability also decreased (Fig. 5 and Table 4). Additionally, correlation analysis between the perceived maneuverability assessed by Q1 through Q4 and the eye movement gain revealed moderate to weak positive correlations with statistical significance (Fig.7-(a) and Table 5). Correlation analysis between the perceived maneuverability assessed by Q1 through Q4 and the eye movement error also revealed weak negative correlations with statistical significance (Fig.7-(b) and Table 6). These findings suggest that differences in maneuverability resulting from changes in machine dynamics are reflected in involuntary eye movements.

### 4.2 Comparison with previous studies

There have been several studies on the relationship between operation in ride-on machinery and involuntary eye movements. For example, it has been shown that the accuracy of involuntary eye movements improves during active motion compared to passive exposure to motion (Uefune et al., 2018).

Furthermore, a study suggests the possibility that this finding can be expanded to differences in the SoA within active conditions (Sato et al., 2020). Specifically, results have suggested that in situations where SoA is significantly diminished—such as when maneuverability is intentionally impaired due to excessive delay or when the correspondence between joystick input and machine rotation is reversed (i.e., the gain becomes negative)—VOR stability decreases. In contrast, a preliminary study    (Sato et al., 2021) examined perceived maneuverability and involuntary eye movements under two distinct machine dynamics—specifically in gain and time constant, while maintaining preserved SoA. This suggests that these relationships may also hold within an active control context. Therefore, the contribution of the present study lies in extending prior work by systematically varying the motion delay (time constant) across multiple conditions while keeping gain fixed, thereby providing a more fine-grained understanding of how subtle variations in machine responsiveness affect both maneuverability perception and involuntary eye movements. A possible interpretation of this series of findings can be found in the observations of (Cullen & Roy, 2004), which suggest that predictive motion information plays an important role in maintaining stable gaze, as motion prediction is also crucial in machinery operation. Furthermore, based on the internal model hypothesis (Kawato, 1999; Merfeld et al., 1999; Wolpert et al., 1995), motion prediction may be regarded as the degree of understanding of the machine's dynamic properties.

On the other hand, several studies have shown that accuracy of involuntary eye movements deteriorates when a mental workload increases (Kono et al., 2023; Le et al., 2020; Obinata et al., 2008). These studies primarily targeted automobile passengers and drivers without altering the machine dynamics while implementing a mental task. Based on these findings, the decrease in the accuracy of involuntary eye movements observed in the present study can also be interpreted as reflecting an increase in cognitive load when operating a machine with low maneuverability. Given that cognitive load is considered to affect self-motion perception and motion control, both the present study and previous research on cognitive load and eye movements (Kono et al., 2023; Le et al., 2020; Obinata et al., 2008) may have observed changes in motion perception. Furthermore, independent of such interpretations, another possible explanation is that when fixation accuracy is low, visual motion perception becomes more difficult, thereby making the operation more difficult. A more in-depth examination of these various possible interpretations is necessary.

4.3 Limitations

The experiment in the present study was conducted under the condition where the fixation point was fixed to the environment, and participants were instructed to maintain fixation on it at all times. It is an important research question to verify whether the findings obtained in the present study hold under more realistic conditions where gaze and head movements are allowed freely. It is worth noting that in the present study, an operating time of only 90 seconds was uniformly applied across all conditions. Consequently, adaptation was unlikely to occur, and the experiment primarily focused on maneuverability before adaptation. Therefore, this design does not allow for discussions on adaptation processes that may emerge with prolonged machine use. In contrast, research (Sato et al., 2021) reported preliminary experimental results using the same yaw-rotational machine, where participants first adapted to a specific dynamic property and then experienced a sudden change in that property while evaluating involuntary eye movements and subjective maneuverability in both conditions. Their preliminary findings suggested that maneuverability decreased and the accuracy of involuntary eye movements declined following the sudden change in dynamic properties, indicating that the findings from the present study could potentially be extended to maneuverability changes arising from adaptation. Further research in this direction is encouraged. It is an important future research issue to investigate the effect of improved maneuverability adapted through prolonged machine operation on involuntary eye movements.

4.4 Conclusion

   To conclude, the present study provides the first systematic evidence that changes in the accuracy of involuntary eye movements are associated with the perceived maneuverability of ride-on machinery induced by systematic variations in the machine's time constant of the dynamic properties. We observed that an increase in the machine's time constant led to decreases in both perceived maneuverability and eye movement accuracy. These findings underscore the importance of evaluating machine suitability for specific tasks and users, and they highlight the potential benefits of designing machine dynamics and control strategies that can adapt to individual operator capabilities.

**Declarations**

*Competing interests*

The authors have no competing interests to declare.

*Funding*

This study was partially supported by JSPS KAKENHI (grant numbers 21K18308 and 24H00298).

*Ethical approval*

   The experiment was conducted based on the approval of the Ethics Review Committee of the Nara Institute of Science and Technology.

*Consent to participate*

   Informed consent was obtained from all individual participants included in the study.

*Data availability statement*

The generated in this study are available from the corresponding author on reasonable request.